\documentclass[journal]{vgtc}                

\ifpdf
  \pdfoutput=1\relax                   
  \usepackage{graphicx}                
  \usepackage{hyperref}
  \DeclareGraphicsExtensions{.pdf,.png,.jpg,.jpeg} 
\else
  \usepackage{graphicx}                
  \DeclareGraphicsExtensions{.eps}     
\fi%

\graphicspath{{figures/}{pictures/}{images/}{./}} 

\usepackage{microtype}                 
\PassOptionsToPackage{warn}{textcomp}  
\usepackage{textcomp}                  
\usepackage{mathptmx}                  
\usepackage{times}                     
\usepackage{cite}                      
\usepackage{tabu}                      
\usepackage{booktabs}                  
\usepackage[dvipsnames]{xcolor}
\usepackage{flushend}

\onlineid{1084}

\vgtccategory{Research}
\vgtcpapertype{application/design study}

\title{Narvis: 
Authoring Narrative Slideshows for \\Introducing Data Visualization Designs }

\shortauthortitle{Wang \MakeLowercase{\textit{et al.}}: Narvis: Authoring Narrative Slideshows for Introducing Data Visualization Designs}
\vspace{-4mm}
\author{Qianwen Wang, Zhen Li, Siwei Fu, Weiwei Cui, Huamin Qu}

 \authorfooter{
 \item
 Q. Wang, Z. Li, S. Fu, and H. Qu are with Hong Kong University of Science and Technology.
 Email: \{qwangbb, zlidn, sfuaa\}@connect.ust.hk, huamin@cse.ust.hk.
 \item
 W. Cui is with Microsoft Research Asia.
 Email: weiwei.cui@microsoft.com.

 }

\abstract{
  Visual designs can be complex in modern data visualization systems, 
  which poses special challenges for explaining them to the non-experts.
  However, few if any presentation tools  are tailored for this purpose. 
  In this study, we present Narvis, a slideshow authoring tool designed for introducing data visualizations to non-experts. 
  Narvis targets two types of end users: 
  teachers, experts in data visualization who produce tutorials for explaining a data visualization, 
  and students, non-experts who try to understand visualization designs through tutorials. 
  We present an analysis of requirements through close discussions with the two types of end users.
  The resulting considerations guide the design and implementation of Narvis.
  Additionally, to help teachers better organize their introduction slideshows, 
  we specify a data visualization as a hierarchical combination of components, 
  which are automatically detected and extracted by Narvis.
  The teachers craft an introduction slideshow through first organizing these components,
  and then explaining them sequentially.
  A series of templates are provided for adding annotations and animations to improve efficiency during the authoring process.
  We evaluate Narvis through a qualitative analysis of the authoring experience,
  and a preliminary evaluation of the generated slideshows.
} 


\CCScatlist{ 
 \CCScat{K.6.1}{Management of Computing and Information Systems}%
{Project and People Management}{Life Cycle};
 \CCScat{K.7.m}{The Computing Profession}{Miscellaneous}{Ethics}
}

 \teaser{
   \includegraphics[width=\linewidth]{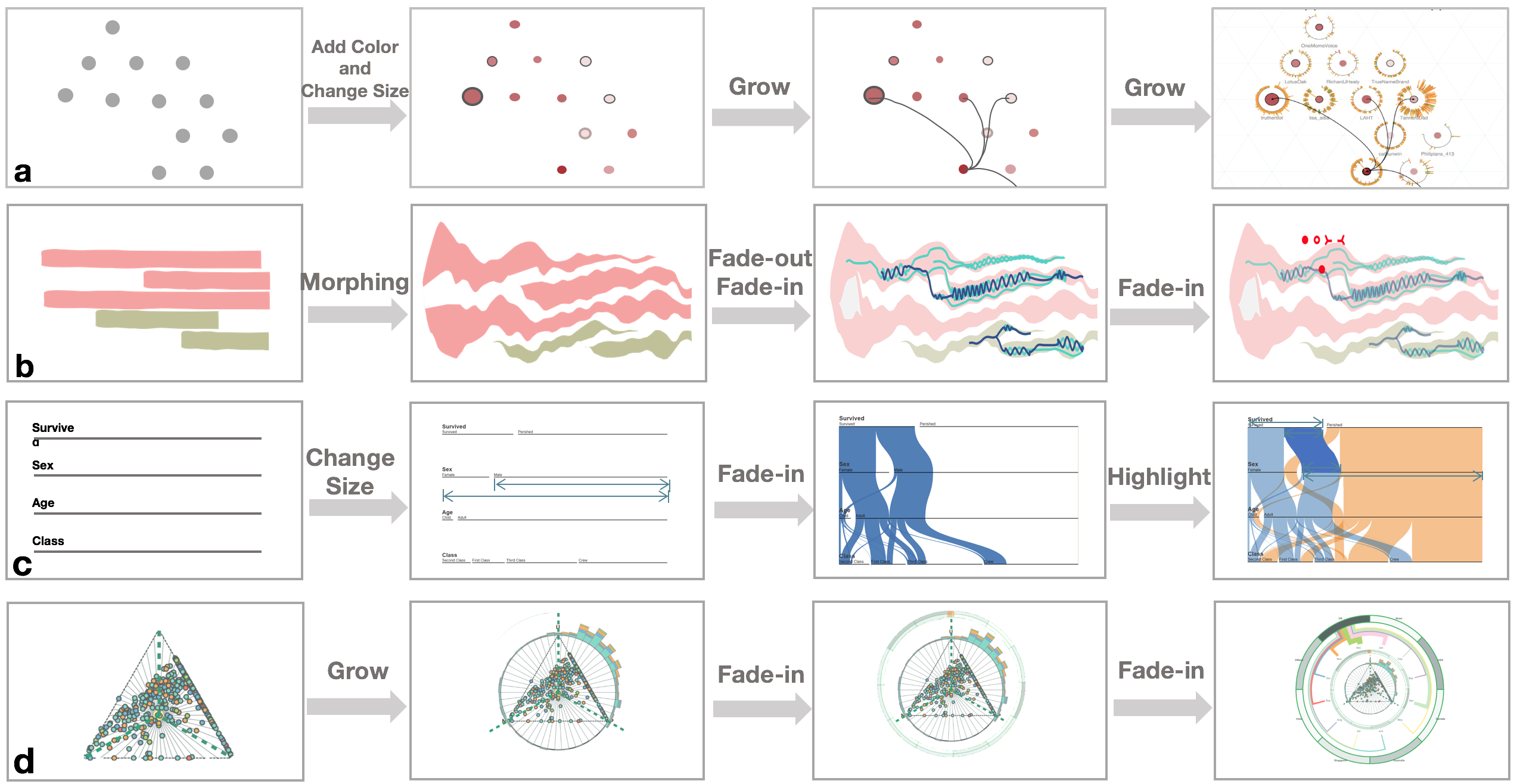}
   \caption{Examples of the slideshows created by Narvis.
     Snapshot images illustrate the content of  four slideshows that introduce four designs, 
     (a) TargetVue~\cite{cao_targetvue:_2016}, (b) Textflow~\cite{cui_textflow:_2011}, (c) Parallel Sets~\cite{ParallerSet}, and (d) OpinionSeer~\cite{wu_opinionseer:_2010}, respectively.
     They demonstrate how different visual channels (\textit{i.e.}, color, size, shape, position) 
     and visual units (\textit{i.e.}, dots, lines, bars, flows) are added sequentially for a progressive introduction.
     These slideshows use seven types of animation templates provided by Narvis: 
     fade-in, fade-out, growing, changing size, adding color, morphing, and highlighting.}
  \label{fig:teaser}
 }

\begin{document}



\maketitle


\section{Introduction}
Data visualization is widely used in data presentation and analysis.
While basic visualization methods (\textit{e.g.,} line charts, bar charts, donut charts) have demonstrated their utility and effectiveness, 
they may fail to present complex data. 
For example, they can hardly give a comprehensive illustration of how the topics in social media split, 
merge and compete with each other.    
  
Over the years, many advanced visualization methods have been devised for communicating and analyzing these complex data. 
However, these advanced visualizations are usually composed of numerous visual components, have diverse visual encodings, and thus are not intuitively understandable (\textit{e.g.,} Fig.~\ref{fig:teaser}(a) and (d)).
Consequently, these advanced visualization techniques haven't been widely exposed to the public,
even though their utility has been verified by experts from various domains~\cite{weber2017apply}.  


Imagine the following scenario:
Jessica, an infovis teacher, wants to introduce a novel visualization technique
described in an academic article in her class. The visual design is complex, as it
is composed of many visual components and various visual encodings that the students likely have never seen before.
She anticipates they might have immediate questions, such as \textit{``what do the links mean?''}, \textit{``what do the positions of nodes mean?''}. So she decides to create a tutorial to guide her students through the different parts of the visualization to help them better understand it.

To create a tutorial, a common solution is to accompany a visual design with a textual description. For example, a whole section is dedicated to describing the visual encodings in the academic article. But learning through an extensive and monotonous textual description is seldom pleasant, and often inefficient. Videos, on the contrary, are relatively engaging, attractive, and comprehensible~\cite{yadav2011if}. Like many other visualization papers, the academic article that Jessica wants to introduce has an accompanying video that demonstrates how to interpret the visual design. However, the pacing and the content of this video cannot be easily edited to fit the needs of the class. Reproducing a video from scratch is generally difficult. Slideshows are a favorable alternative. As with videos, slideshows reveal information progressively in an engaging and appealing manner while being easier to produce. Moreover, the pace of slideshows can be easily controlled by either the teacher or the students.


However, crafting a comprehensible and attractive tutorial slideshow can be challenging.
First, to make the slideshow attractive and engaging, some teachers may apply graphic editing techniques such as ``\textit{highlighting}'', ``\textit{morphing}'' and ``\textit{zooming in}''.
However, these techniques can be tedious and time-consuming to achieve using existing tools.
Second, due to \textit{``the curse of knowledge''}~\cite{wiemanncurse}, 
these teachers might unknowingly assume that their students have the background to understand and unconsciously omit the explanation for certain visual encodings and components. 
Third, collecting feedback from the students can also be difficult for teachers, which directly hinders the teachers from improving the quality of their tutorials. 

Moreover,  it is also challenging to make the slideshow tutorial comprehensible to the students.
First, the students can easily get overwhelmed if they are inundated with all the information simultaneously.
At the same time, an improper narrative sequence will confuse the students considering the logic dependence existing among visual components.
For example, nodes in a node-link diagram should be introduced before the links connecting them. 
In an advanced visualization design, which has more components than merely nodes and links, identifying a proper explanation sequence can be more challenging and demanding.
In addition, the students can be easily distracted and become lost, thereby preventing effective learning.


To tackle these challenges, we propose Narvis, 
a slideshow authoring tool 
that assists in introducing advanced visualization designs. 
Those who craft tutorials are the direct end users of Narvis.
Meanwhile, it is important to take those who read these tutorials into account when designing a narrative data visualization ~\cite{riche2018data}.
Thus, two types of end users are identified for Narvis: teachers and students. 
We conduct interviews with these two groups of end users to understand their expectations, identify the problems in existing tools, 
and then distill the design requirements for Narvis.
At the same time, to form a clear logic in the tutorial slideshows,
we borrow lessons from previous work about the design space of a data visualization~\cite{munzner_visualization_2014, von2002language}, 
combine them with our empirical observations, 
and propose an approach to decompose a visualization design into several components and to introduce these components progressively. 
Based on the aforementioned analysis, we design and implement Narvis, 
aiming to offer an efficient, expressive and user-friendly authoring tool for introducing data visualization designs. 
We evaluate Narvis in terms of authoring experiences and the quality of generated slideshows.

Our contributions are as follows: 
1) An analysis of the requirements for an authoring tool tailored for introducing a visualization design.
These requirements are obtained from the perspectives of teachers and students. 
2) An approach to hierarchically decompose a visualization design and introduce its compositions progressively.
3) The design and implementation of Narvis, a prototype authoring tool to generate and edit tutorial slideshows for introducing visualization designs.
We believe our work can facilitate wider adoption of advanced visualization designs.

\section {Related Work}\label{sec:related_work}
In this section, we discuss the prior studies that are closely related to our work.

\subsection{Structures of Narrative Data Visualizations}
Narrative is an effective tool for information communication~\cite{cunningham_culture_2009} and has been widely studied in the fields of literature, comics, and cinema~\cite{shen2011living, cohn_visual_2013}. 
With the increasing importance of data communication, researchers in the visualization community are studying how narrative can be used in data visualization~\cite{wang_animated_2016, riche2018data, boy2015storytelling, mckenna2017visual}.

Some researchers borrowed the concepts from other fields to guide the design of narrative data visualization.
For instance, Amini et al.~\cite{amini_understanding_2015} borrowed concepts about narrative categories from comics to analyze the structure of narrative data visualizations. 
Wang et al.~\cite{wang_animated_2016} adopt two representative tactics, time-remapping and foreshadowing,
from cinematographers to organize structures of narrative data visualization for better information communication. 
Some researchers focused on the narrative structures exclusively for data visualization. 
Satyanarayan and Heer~\cite{satyanarayan_authoring_2014}, through interviews with professional journalists, 
defined the core abstractions of narrative data visualization as state-based scenes, visualization parameters, dynamic graphical and textual annotations, and interaction triggers. 
By identifying the change in data attributes, 
Hullman et al.~\cite{hullman_deeper_2013} proposed a graph-driven approach to automatically identify effective narrative sequences for linearly presenting a set of visualizations. 
These works, however, mainly focus on conveying the insight discovering process and rarely discuss the narrative structures used for explaining visualization designs.

Maybe the closest to our work is the narrative introduction proposed in the work of Boy et al.~\cite{boy2015storytelling}, which provides initial insights and introduces different visual
encodings for three online data visualizations.
The authors found that these narrative introductions, even though increased uptime and number of visits, didn't increase user-engagement in exploration. 
However, they focused on studying the effect of narrative introductions on user-engagement and paid little attention to analyzing the proper structure of a narrative introduction.

Compared with previous studies, we aim to help people identify a proper narrative sequence for introducing the visual encoding of a visualization design.

\subsection{Decomposing Data Visualizations}
The clarification of the design space of a data visualization design can help people get a better understanding of this design. 
Munzner~\cite{munzner_visualization_2014} proposed that a data visualization design 
\textit{``can be described as an orthogonal combination of two aspects: graphical elements called marks and visual channels to control their appearance''}. 
Borrowing the concept of physical building blocks such as Lego, 
Huron et al.~\cite{huron_constructive_2014} extended the design space of a data visualization, 
defining the components of a data visualization as a token, token grammar, environment and assembly model.
Javed et al.~\cite{javed2012exploring} focused on a high-level structure, exploring the composition of multiple visualization views.
These theoretical works motivate the designers of visualization tools to develop efficient high-level visualization systems~\cite{bostock_protovis:_2009,mendez_ivolver:_2016}.  

On the other hand, theoretically identifying the basic components of a data visualization enables people to physically extract them. 
Harper and Agrawala~\cite{harper_deconstructing_2014} contributed a tool that extracts visual variables from existing online visualization designs to generate a new design. 
Huang et al.~\cite{Huang:2007:SUI:1284420.1284427} proposed a system that recognizes and interprets imaged infographics from a scanned document. 
ReVision~\cite{savva_revision:_2011} applied computer vision methods to recognize the types, marks, 
encodings of a data visualization, and allows the users to create a new design based on these data. 

Narvis is inspired by previous theoretical analysis. However, instead of decomposing a current design and mapping it to an alternative one, 
Narvis extracts the basic compositions of a visual design for introducing this design progressively and structurally.



\subsection{Authoring Tools for Narrative Visualizations}
The extensive requirements of data communication have motivated researchers to investigate techniques for authoring narrative visualization. 

User experience represents a vital concern when utilizing an authoring tool. 
SketchStory~\cite{lee_sketchstory:_2013}, with its free-form sketching of interaction, provides an engaging approach to creating and presenting narrative visualizations. 
DataClips~\cite{amini_authoring_2017} lowers the barrier of crafting a narrative visualization by providing a library of data clips, 
thereby allowing non-experts to be involved in producing narrative visualizations. 
Lyra\cite{2014-lyra} allows designers to author expressive visualization designs through drag-and-drop interactions without writing code.

Information delivery is the core consideration of an authoring tool. 
Existing authoring tools typically select a specific type of narrative visualization based on the information type~\cite{amini_authoring_2017, fulda_timelinecurator:_2016}. 
Meanwhile, integrating an authoring tool for narrative visualization with a  data analysis tool is becoming a trend 
since it effectively bridges the gap between data analysis and communication~\cite{eccles_stories_2007, bryan_temporal_2016,lee_more_2015}. 
 
These tools offer inspiring user interaction design and provide favorable examples for implementing narrative visualization. 
However, these tools treat visual encoding as a cognitively obvious attribute that can be universally recognized without a formal introduction.
Thus, these tools are ill-suited for introducing visualization designs. 
Contrary to the aforementioned tools, Narvis enables people to produce narrative tutorial slideshows, which introduce visualization designs by decomposing a visualization design and explaining their visual components progressively
.
\section{Understanding End Users}\label{sec:design_requirement}
To better guide the design and implementation of Narvis, 
we conducted interviews with two groups of end users (\textit{i.e.}, teachers and students) to investigate their current practice of producing tutorials and understanding visualizations. 
Six design requirements for Narvis were derived from our interviews.

\subsection{Users}


To investigate the current practice of making tutorials and the experience of reading tutorials, 
we collaborated with the two types of end users of Narvis, \textit{i.e.,} teachers who craft tutorials and students who read tutorials. 
In this study, we assume that teachers have extensive experience in visualization designs and students have little prior knowledge of data visualization.
The editor group consisted of two teaching assistants (TAs) in a data visualization course, 
one professor in the field of data visualization, 
and two employees from a commercial data visualization website.
All the five teachers, denoted as T1 to T5, have the needs to produce tutorials for introducing visualization designs.
A slideshow is their preferred type because it is \textit{``more appealing and comprehensible than textual descriptions and easy to edit than videos''} (T2).


The students group consisted of seven undergraduate students, denoted as S1 to S7, from three departments (\textit{i.e.}, biology, finance, computer science). 
All these students had less than three-months experience in data visualization. 
They recently took a data visualization course, where they were required to survey visual designs for their assignments. 
Thus they learned many visual designs from the internet and watched many tutorials.

\subsection{Interviews}
We conducted two-part, semi-structured interviews with teachers and students separately. 

In the first part, each teacher described their recent experience in producing an introduction slideshow.
We encouraged them to open this slideshow and recall as many details as possible.
Then, we inquired more about the organization of the narrative sequence in their tutorials, such as \textit{``why the encoding of color is introduced before the encoding of position?''} 
In the second part, teachers enumerated all the tools they have ever used, 
identified the most frequently-used tool, and provided their reasons for their choice.
Then, we asked them about the obstacles they encountered when using this tool, 
and how they overcame these obstacles.
The entire interview lasted approximately 40 minutes for each participant.

In the interview with the students, we discussed with them the tutorials they had learned from.
We first asked their general comments about the different types of tutorials (\textit{e.g.,} online blogs, course lectures, papers, blogs). 
Then each student identified one positive example and one negative example they encountered.

These tutorials are mainly about how to read a visualization design as a combination of graphics.
For example, they regarded the encoding explanation (0:52-1:35) in a video \footnote{https://www.youtube.com/watch?v=XQ6xPkAZsPU}
and the encoding explanation in a blog\footnote{http://www.dear-data.com/week-08-1/} as positive examples that are \textit{``comprehensible and helpful''}.

We discussed example tutorials with them, listened to their comments, 
and identified the obstacles toward their understanding.
The length of the interviews varied from person to person, lasting from 18  to 43 minutes.

We took notes and videos during interviews for later analysis. 

\subsection{Design Requirements}

Narvis aims to help teachers produce narrative slideshows for introducing data visualization designs.
Thus, support for common operations and guidance to avoid common mistakes should be provided in Narvis.

Based on our observations during the interviews, we categorized six design considerations for Narvis, denoted as R1 to R6. 


\noindent
\textbf{R1. Enable Efficient and Expressive Graphic Editing.}
T1, T2 and T3 (two TAs and one professor) typically used generalized presentation tools (\textit{e.g.,} PowerPoint, Keynote, Prezi) because of their high efficiency,
although the graphic editing capacities of these tools are limited.
\textit{``Highlighting a subarea is effective while introducing a visual design, yet it can hardly be achieved in PowerPoint''}, commented by T1.
Owing to the complexity of the operation, professional graphic editing software (\textit{e.g.,} Illustrator, Photoshop) were only used for special cases, 
such as \textit{ ``a demo influencing the next year's funding''} (T3, the professor). 
T4 and T5, two employees from an online data visualization platform, preferred to use professional software.
But the producing process was time-consuming, taking nearly \textit{``two weeks to produce one tutorial''}.
A gap remains between the efficiency of general tools and expressiveness of professional graphical tools.

\noindent
\textbf{R2. Avoid Unconscious Overlooking.}
T1, T2, and T4 all mentioned that they might unconsciously miss the explanation of certain visual encodings without a good preparation.
Since an advanced visualization design usually consists of various visual components with varying visual encodings, it can be challenging to ensure that every visual component and every visual encoding are properly explained.
Moreover, with extensive expertise in data visualization, teachers might treat certain visual encodings as self-evident and offer no additional explanation.
However, the incompleteness of information impedes the quality of produced tutorials.

This phenomenon is also known as \textit{``the curse of knowledge''}~\cite{wiemanncurse}. 
Well-informed agents (\textit{e.g.,} experts in data visualization)
typically assume the less-informed agents (\textit{e.g.,} students with little prior knowledge of data visualization) have the background to understand. 


\noindent
\textbf{R3. Collect Feedback.} 
For teachers, communicating with their students can be difficult, 
especially when the student size is large 
or when the student is remote in time and space (\textit{e.g.}, the student of an online course).
\textit{``It is hard to make sure that all people understand all the visual encodings correctly. 
The thought that they may interpret the data falsely always bothers me''}, remarked T4, an employee from an online data visualization platform.

However, teachers require student feedback to evaluate and improve their tutorials.
All interviewees in the teachers group stated that they would show their produced slideshows to their friends or colleagues for quality evaluation and further revision.
But \textit{``such evaluation methods can be biased since my friends are already familiar with visual designs''} (T5),
and \textit{``it is hard to collect feedback in a wide range''} (T2).

\noindent
\textbf{R4. Avoid information overload.} 
Six out of seven participants in the students group complained that they experienced information overload when reading tutorials,
such as a paper that \textit{``uses one even two pages to describe a visualization design''}
or slideshows that \textit{``put too many things in one slide''}.
S3 stated that \textit{``I just skip some parts when I am inundated with too much information.''} 

Complex visual designs contain numerous visual components with varying visual encodings, 
thus imposing a cognitive burden on the students.
To make a tutorial slideshow comprehensible, these visual components should be introduced progressively to avoid overloading the students.

\noindent
\textbf{R5. Provide Clear Narrative Logic.}
The interview with students revealed that many web-based data visualization systems and visualization designs were rarely accompanied by detailed, comprehensive tutorials.
Students complained a lot about the lack of clear logic in these tutorials.
\textit{``Sometimes, I had to read a tutorial several times, reorganize all the information myself to fully understand a visual design''}, stated by S4.
When commenting on a tutorial, S2 said, 
\textit{``It first explains `A', then it explains `B', and suddenly it goes back to explaining `A' again. Maybe it has its own logic, but I am totally confused.''}

Creating a tutorial slideshow with clear logic and informing students of this logic will facilitate their understanding of a visualization design.


\noindent
\textbf{R6. Keep the Sense of Overview.}
When receiving a large amount of information (\textit{e.g.,} a visualization design contains many visual components with varying visual encodings), 
the students can be easily distracted or forget previous information.
In this situation, informing the students of the overall structure of a visualization design and reminding them of the previous messages can aid the perception process. 

\begin{figure}
 \centering 
 \includegraphics[width=\columnwidth]{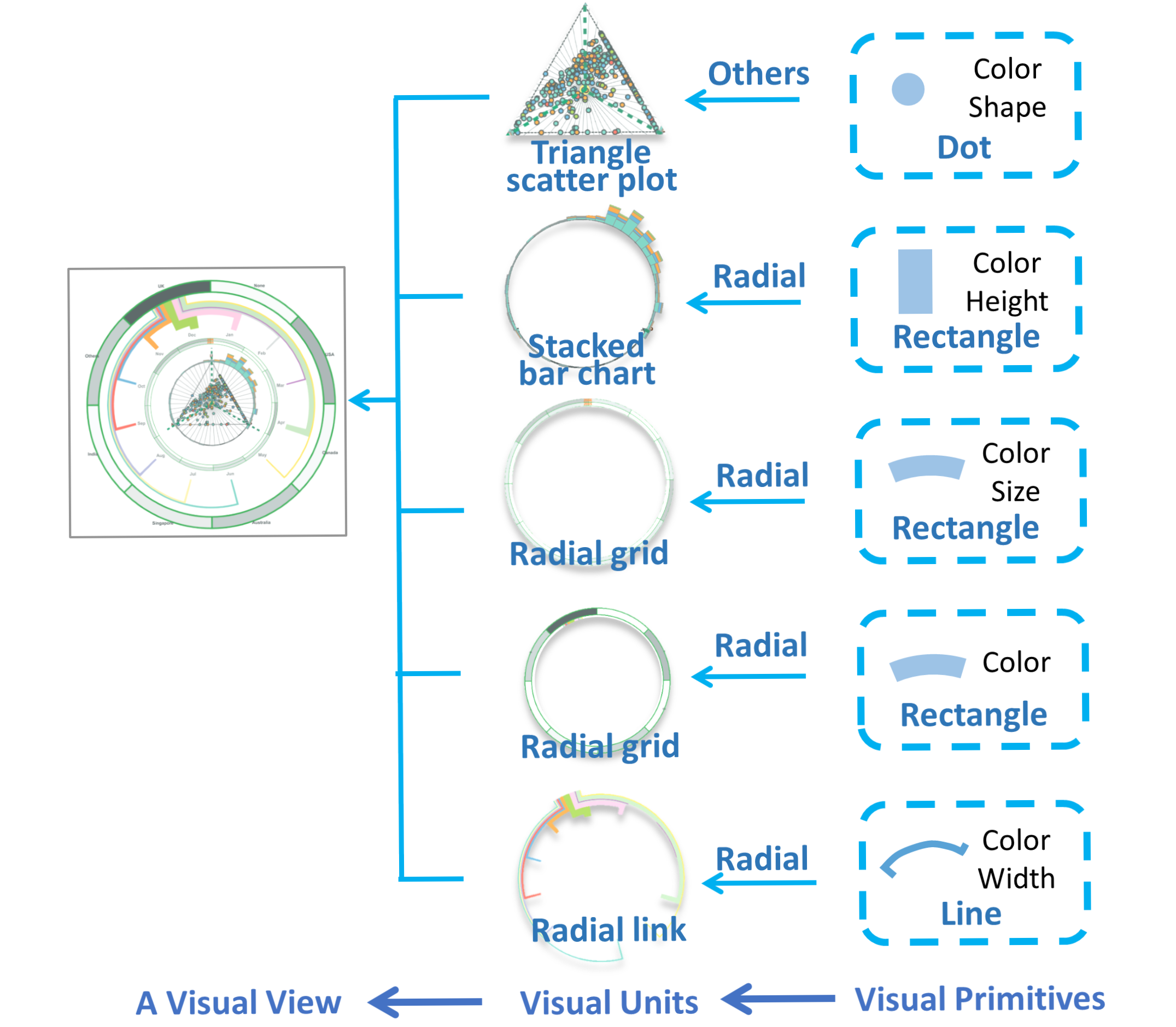}
 \caption{After decomposing OpinionSeer\cite{wu_opinionseer:_2010}, we obtain the hierarchical structure of this visualization. It consists of five visual units and employs three kinds of visual primitives. }
 \label{fig:hierarchic}
\end{figure}

\section{Understanding Data Visualization} \label{sec:analysis}
According to the aforementioned design requirements, 
we realize that a comprehensive understanding of data visualizations is crucial for designing and implementing Narvis.
Only by understanding a visualization design 
can Narvis know how to guide the teachers to form a clear narrative sequence (R2),
assist them in making a progressive introduction (R4),
prevent them from omitting certain visual encodings (R5),
and help them keep the students concentrated (R6).




In this section, we aim to understand a visualization design by answering the following three questions: 
``\textit{What are the basic components that compose a data visualization? }'' ,
``\textit{What is the relationship between these components? }'', 
``\textit{How should we deal with these relationships when introducing a data visualization? }''

\subsection{Components of a Visualization}\label{subsec:compositions}
Efforts have been made to identify the atomic building blocks of a visualization~\cite{mendez_ivolver:_2016, bertin1983semiology, von2002language}. 
In this study, we extend the previous work by 
1) proposing a hierarchical structure for decomposing a visual design, 
2) including the logic relationship between components. 
Here, we define a visualization as a single-view, static presentation of data.

In our model, a visualization is a hierarchical structure of three levels, namely,
visual primitives, visual units, and visual views.
Taking OpinionSeer~\cite{wu_opinionseer:_2010} as an example, 
we apply the hierarchical model and decompose this design into five visual units, 
as depicted in Fig.~\ref{fig:hierarchic}. 

\textbf{A Visual Primitive} 
is a graphical element whose visual channels, such as color, width, and height, are mapped to data attributes with certain visual grammars. 
Visual channels are visual properties that control the appearance of a graphical element, 
whereas a visual grammar describes the way a visual channel represents a data attribute. 
For instance, a point is a visual primitive, size is a visual channel, and \textit{``size indicates the importance score''} is a visual grammar. 

\textbf{A Visual Unit} is an assembly of visual primitives that are bound with the same data attributes.
For example, one dot is a visual primitive, while the dots in a scatter plot constitute a visual unit.
A visual unit is the smallest functional unit of a visualization. 
A visual primitive alone (\textit{e.g.,} a point) has no meaning, and it only has meaning as a part of a visual unit (\textit{e.g.,} an outlier point in a scatter plot). 

\textbf{A Visual View} can be considered a combination of visual units. 
A simple visual view contains only one visual unit (\textit{e.g.,} the lines in a line chart) 
whereas an advanced visual view typically combines multiple visual units.
For example, OpinionSeer ~\cite{wu_opinionseer:_2010} is composed of five visual units, as illustrated in Fig.~\ref{fig:hierarchic}. 

  

\begin{figure}
 \centering 
 \includegraphics[width=\columnwidth]{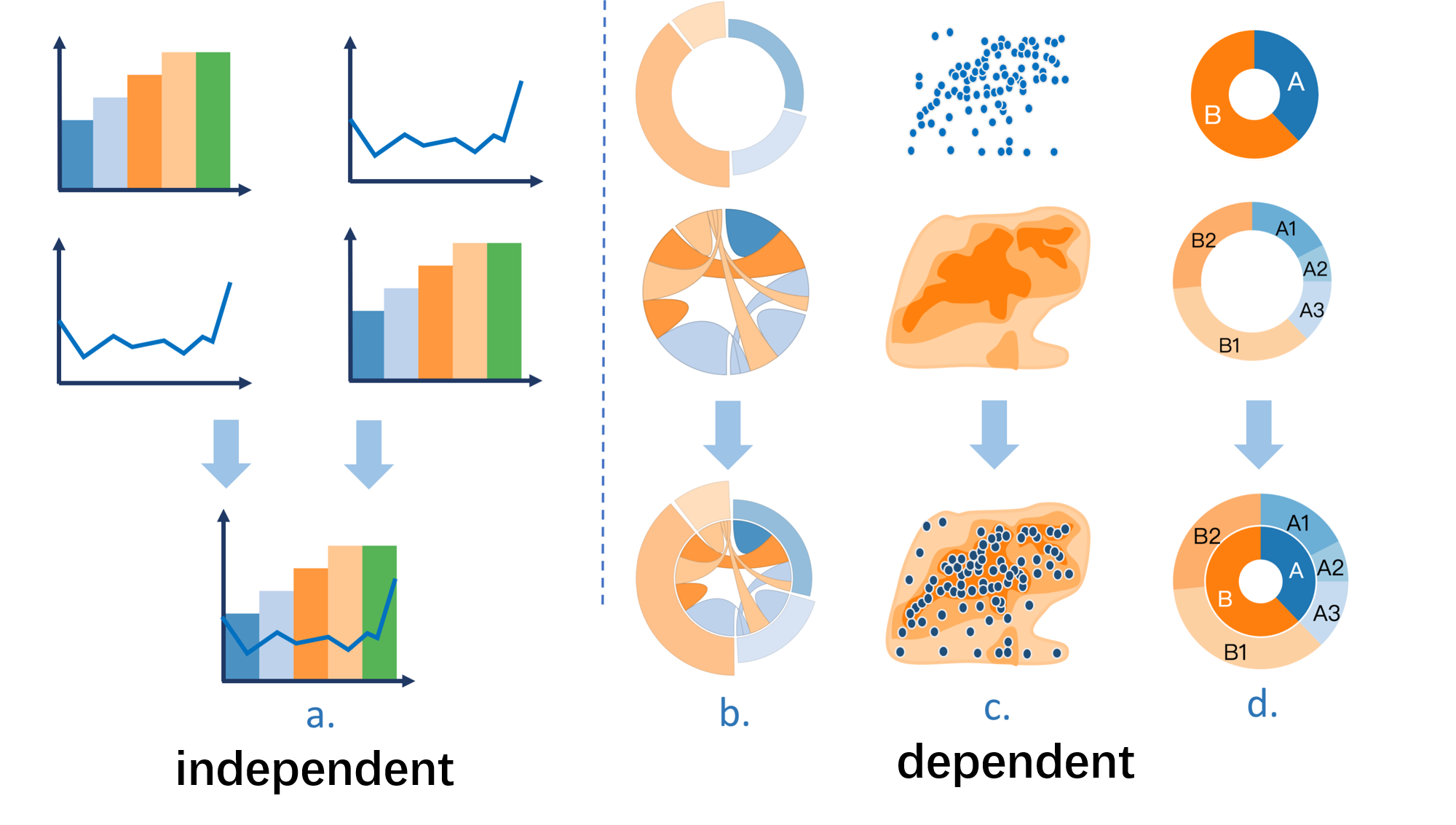}
 \caption{Each visualization on the bottom is comprised of two visual units.
 The relationship between these two visual units can be independent (a), which means that the explanation order of these two units can be arbitrary, or dependent (b, c, and d), which means that one unit must be explained before another. }
 \label{fig:unit_relationship}
\end{figure}

\begin{figure*}[t]
\centering 
\includegraphics[width=\linewidth]{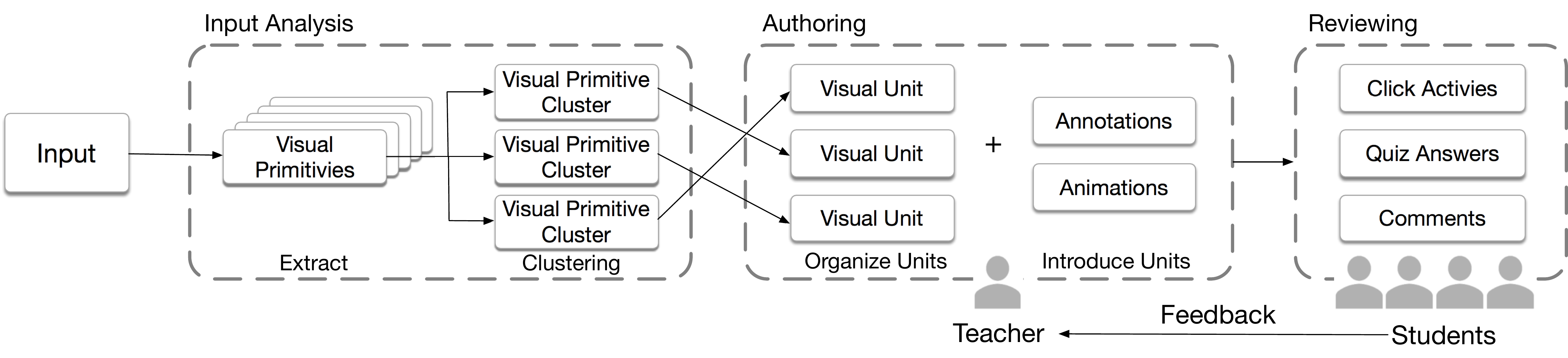}
\caption{Abstract representation of the workflow of Narvis, which consists of Input Analysis, Authoring, Reviewing. 
Two types of end users, namely, teachers and students, are involved.
In Input Analysis, Narvis automatically detects visual primitives from the input and clusters them. In Authoring, teachers craft an introduction slideshow by organizing visual units and explaining them through animated transitions and annotations.
In Reviewing, students watch the produced slideshow and offer feedback, which help the teachers refine the slideshows}
\label{fig:overview}
\end{figure*}

\begin{figure*}[t]
\centering 
\includegraphics[width=\linewidth]{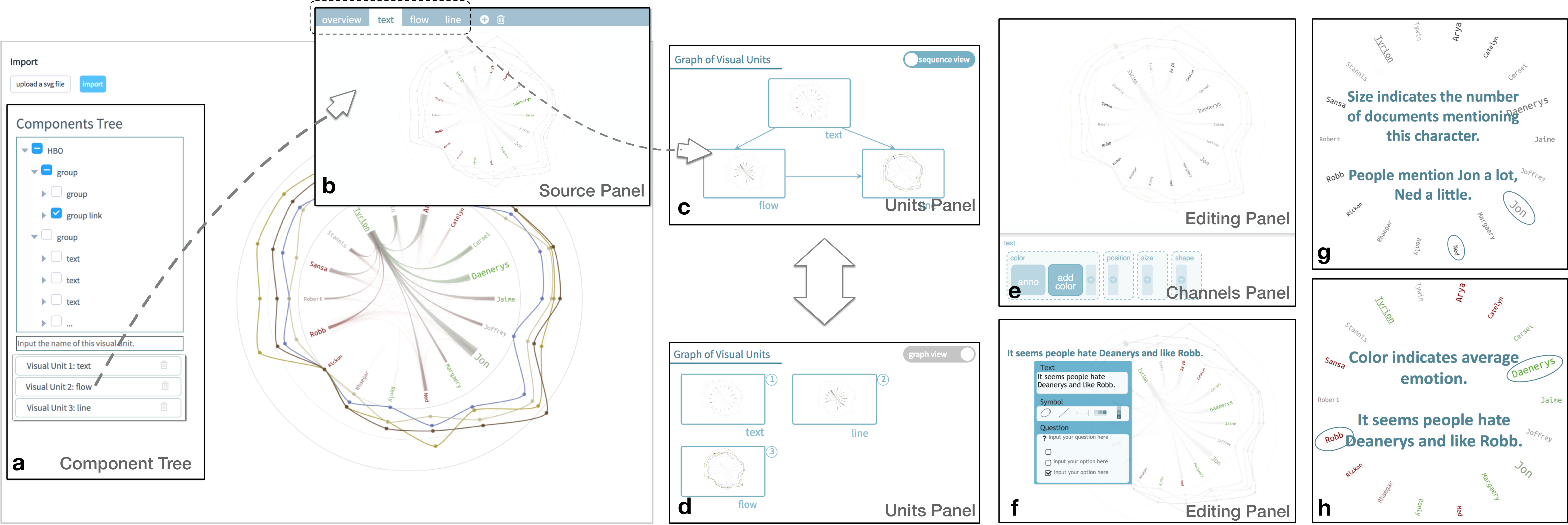}
\caption{The workflow of creating a slideshow with Narvis to introduce the visualization proposed by Scharl et al~\cite{scharl2016analyzing}.
Teachers select visual units from Component Tree (a), identify the logical relationship between viusal units (b) and get a narrative sequence (c). 
They then add animated transitions (e) and annotations (f) for explaining the visual encodings in these visual units.
(g) and (h) are examples of produced slides.}

\label{fig:GOT}
\end{figure*}

\subsection{Relationships between Components}\label{relationship}
We first describe the relationships between conceptual components, 
and then offer suggestions for constructing a narrative sequence based on these relationships. 

\subsubsection{Relationships between Visual Channels}\label{subsubsec: rel_channels}
For a visual primitive, various channels are encoded with different data attributes. 
Thus, these visual channels usually have no logical dependency between themselves. Determining a narrative sequence from their inner logical dependency is difficult. 

Therefore, we define two metrics to arrange visual channels: 
\textbf{the complexity of their encoded information} and \textbf{saliency of their visual appearance}.
In this study, ``saliency'' represents the degree of difficulty for people to notice a certain channel.
The visual saliency of different channels is relatively constant and well-defined ~\cite{munzner_visualization_2014,cleveland_graphical_1984}.

These two metrics are used for the following reasons.
First, the order of decreasing visual saliency can facilitate graphical perception~\cite{cleveland_graphical_1984}. 
Different channels have intrinsically different perceptual saliences and channel with high salience will suppress the expression of others.
This salience strength can be influenced in a task-dependent manner ~\cite{nothdurft_salience_2000}. 
After introducing the channel with high saliency first, we remove this channel from the task list in our mind~\cite{itti2001computational}, 
decrease its saliency and allow other channels an opportunity to attract the limited human attention. 

Second, the order of increasing complexity leads to an effective learning process. 
The easy-to-difficulty procedure has been confirmed as effective for learning new tasks~\cite{bliss_effects_1992}.

\subsubsection{Relationships between Visual Primitives}
Visual Primitives assemble to form a visual unit by following different construction rules.
For example, dots can constitute a scatter plot, a spiral dot chart, or a circle packing chart by following radial, orthogonal, or metric-based construction rules, respectively~\cite{kucher2015text}. 

\subsubsection{Relationships between Visual Units}\label{subsubsec:rel_units}
A visual view can be specified as the combination of several visual units. 
We identify two types of relationships between visual units: dependent and independent relationships. 

\noindent
\textbf{Independent relationship} refers to a relationship that no logic dependency exists between two visual units. 
The two visual units should be explained together but the sequence can be arbitrary. 
For instance, a unit of ``line'', which indicates the temperature over a time period, 
and a unit of ``bar'', which denotes the precipitation over a time period, are jointly placed in one visual view, sharing the same x-axis (Fig.~\ref{fig:unit_relationship}(a)).
The relationship between the two units is independent.
In this situation, the unit ``lines'' can be introduced before or after another unit without impeding understanding.

\noindent
\textbf{Dependent relationship} refers to a relationship wherein one visual unit ``A'' depends on another visual unit ``B''.  
The understanding of visual unit ``B'' is the prerequisite of understanding ``A'', thus ``B'' should be explained before ``A''. 
For example, in Fig.~\ref{fig:unit_relationship} (b), 
a unit of ``flows'' represents the correlation between ``bars''. 
The students need to understand the meaning of ``bars''; thus, flows among these ``bars'' can be meaningful.

\section{Narvis:  A Slideshow Authoring tool}


Guided by the design requirements discussed in Section~\ref{sec:design_requirement}, 
as well as the theoretical model discussed in Section~\ref{sec:analysis}, 
we design and implement Narvis, a slideshow authoring tool for the introduction of data visualization designs. 
The workflow of Narvis consists of three phases (Fig.~\ref{fig:overview}), \textit{i.e.,} Input Analysis Phase, Authoring Phase, and Viewing Phase.

\begin{figure}[tb]
\centering 
\includegraphics[width=\columnwidth]{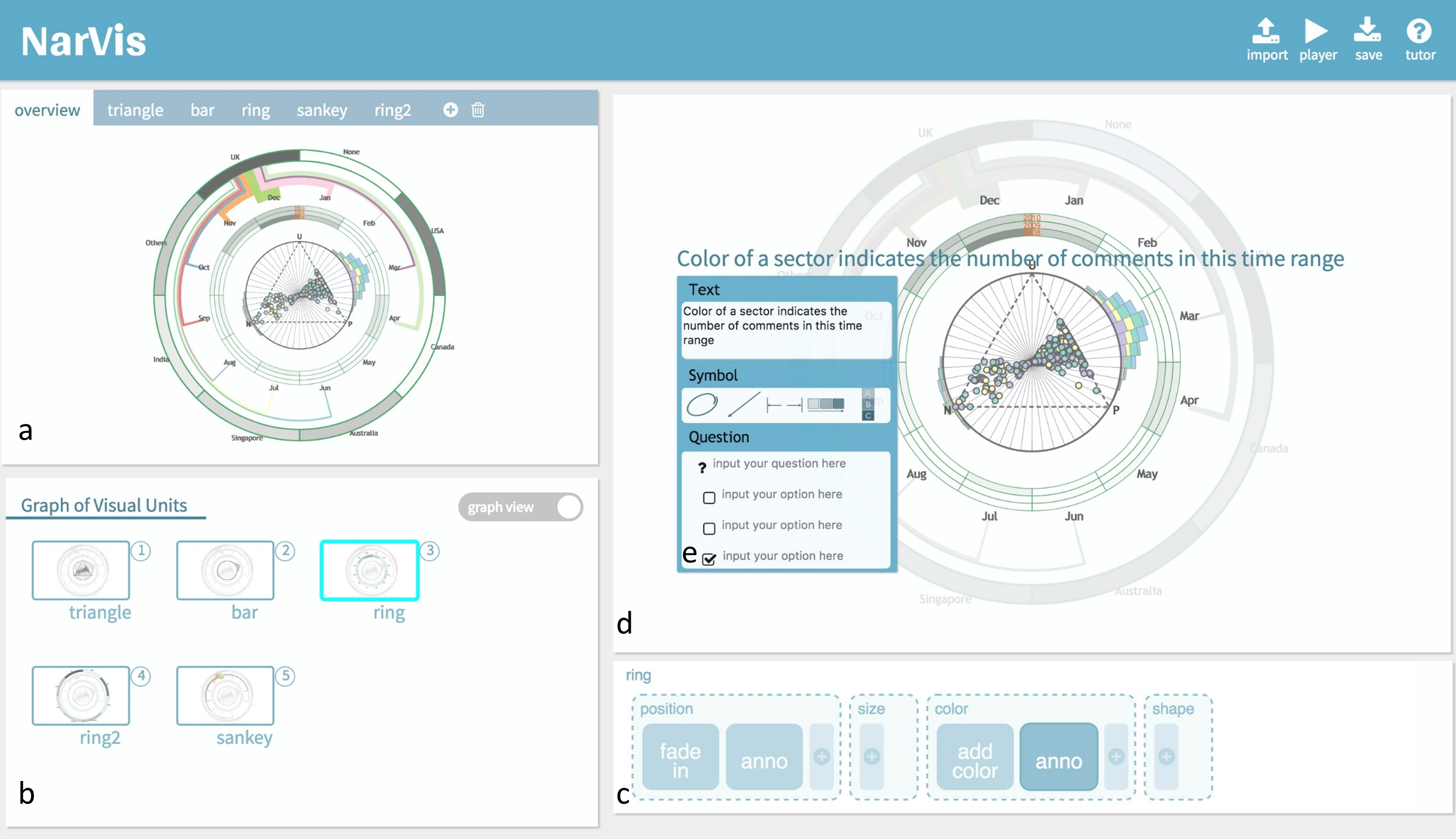}
\caption{Annotated screenshot of the interface of Narvis: 
a) Source Panel where Narvis process the input data, 
b) Units Panel where teachers define the relationships between units and obtain a narrative sequence, 
c) Channels Panel where teachers add annotations and animated transitions for the explanation of each unit, 
d) floating annotation window that offers options for adding annotations,
e) Editing Panel where teachers modify the added animations and annotations.}
\label{fig:interface}
\end{figure}  

\begin{figure}[tb]
\centering 
\includegraphics[width=\columnwidth]{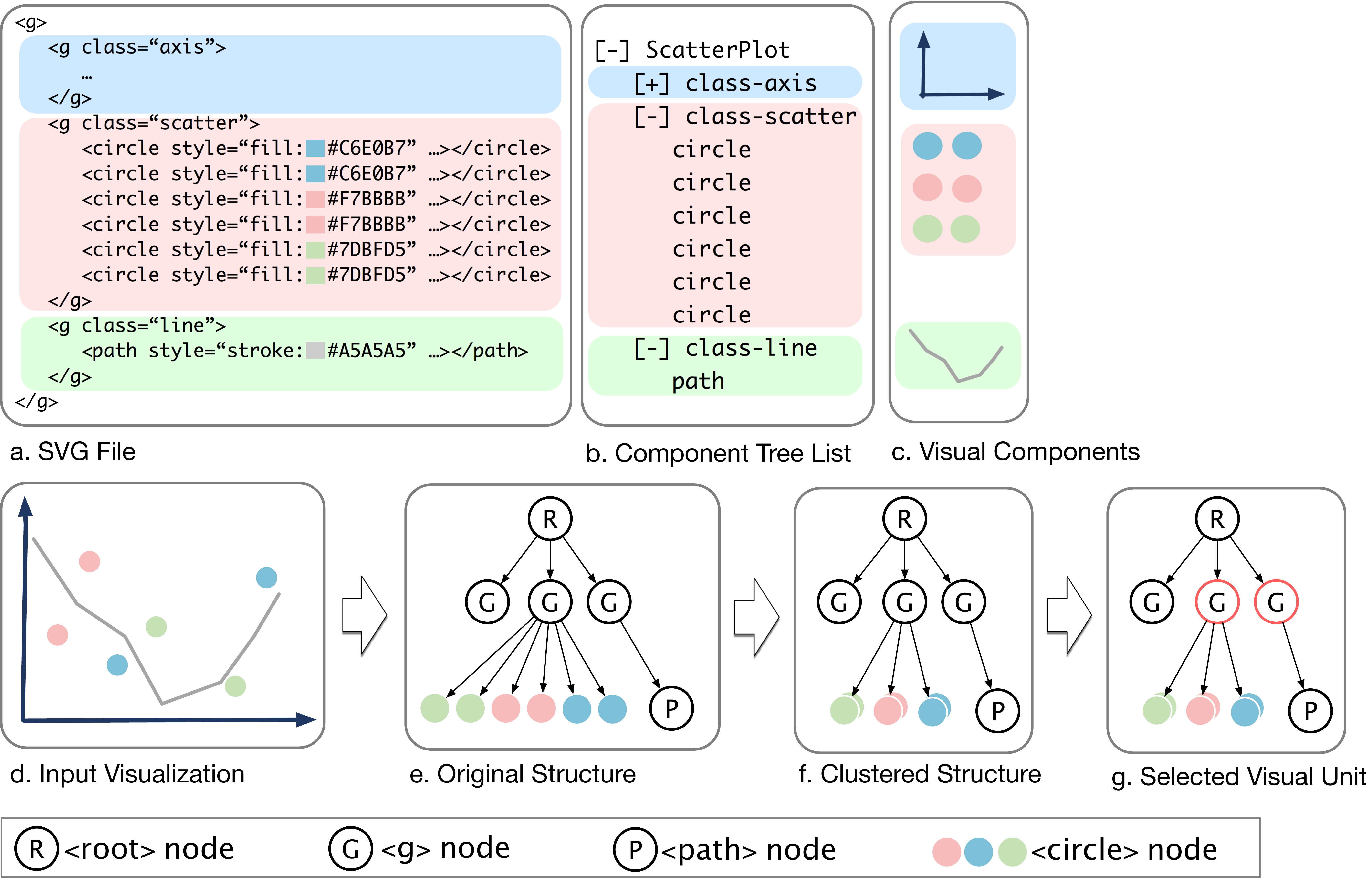}
\caption{Narvis automatically detects and clusters the graphic elements (c) in an SVG file (a) and oragnizes them in the form of an SVG tree (d-g).
Teachers can examine, select, and organize needed visual units from a checkable list (b). }
\label{fig:svgtree}
\end{figure}

\subsection{Phase1: Input Analysis}
In the Input Analysis Phase, Narvis processes a visualization design, extracts the graphic elements from the visual design, and classifies these elements into groups for further editing, as presented in Fig.~\ref{fig:svgtree}. 
  
Narvis uses a scalable vector graphics (SVG) file as input 
because SVG is employed in a wide range of data visualizations, and 
provides a complete scene graph for describing two-dimensional based vector graphics. 
A SVG file describes a scene graph through two types of elements: 
a) shape elements such as $<circle>$, $<rect>$, and $<path>$,
which create basic shape elements on screen;
b) group elements $<g>$ which are the containers used
to group shape elements. 
Each shape element has attributes (\textit{e.g.,} fill, stroke) that define its visual appearance and 
attributes (\textit{e.g.,} class, id) that describe its function. 


Narvis processes a SVG file by extracting its shape elements  and their attributes.
Since  shape elements are essential components that define a SVG,
Narvis is able to parse any  SVG file regardless of its generator.
The extracted shape elements are designated as visual primitives.
They are grouped based on, firstly, their original groups (\textit{i.e.,} $<g>$) and classes, secondly, their element types (\textit{e.g.,} $<circle>, <rect>$), and thirdly, their visual appearance (\textit{e.g.,} color).
It's worth noting that original groups and classes don't necessarily exist in a SVG file.
After clustering, we get a hierarchical structure of these visual primitives.

To identify the visual units in a visualization design, teachers manipulate on the tree list in Components Tree View, as shown in Fig.~\ref{fig:GOT}(a). 

This tree list depicts extracted visual primitives and their hierarchical structures (Fig.~\ref{fig:svgtree}).
Teachers identify visual units by selecting root nodes of subtrees in the tree list. 
All leaf nodes (\textit{i.e.,} visual primitives) of a selected subtree then form a visual unit.
In case teachers are not satisfied with the automatic clustering results, they are allowed to modify this tree list, including split, merge, and remove nodes.
Besides, hovering over a node in the tree list will highlight all visual primitives that are descendants  of this node, which displays at the right side of the tree list (Fig.~\ref{fig:GOT}(a)).

These visual units are displayed at the \textbf{Source Panel}, 
where each tabbed panel contains a cluster of visual primitives and acts as a visual unit (Fig.~\ref{fig:GOT}(b)).
Each tabbed panel can be renamed by the teachers to facilitate further authoring process.

Through input analysis, teachers identify the visual units of the input visualization design.
Graphic elements belonging to the same visual unit are bound together and can be edited together during explanation (R1).
Thus, Narvis allows the teachers an efficient and structural manipulation in the succeeding Authoring Phase. 


\begin{figure}[tb]
\centering 
\includegraphics[width=0.8\columnwidth]{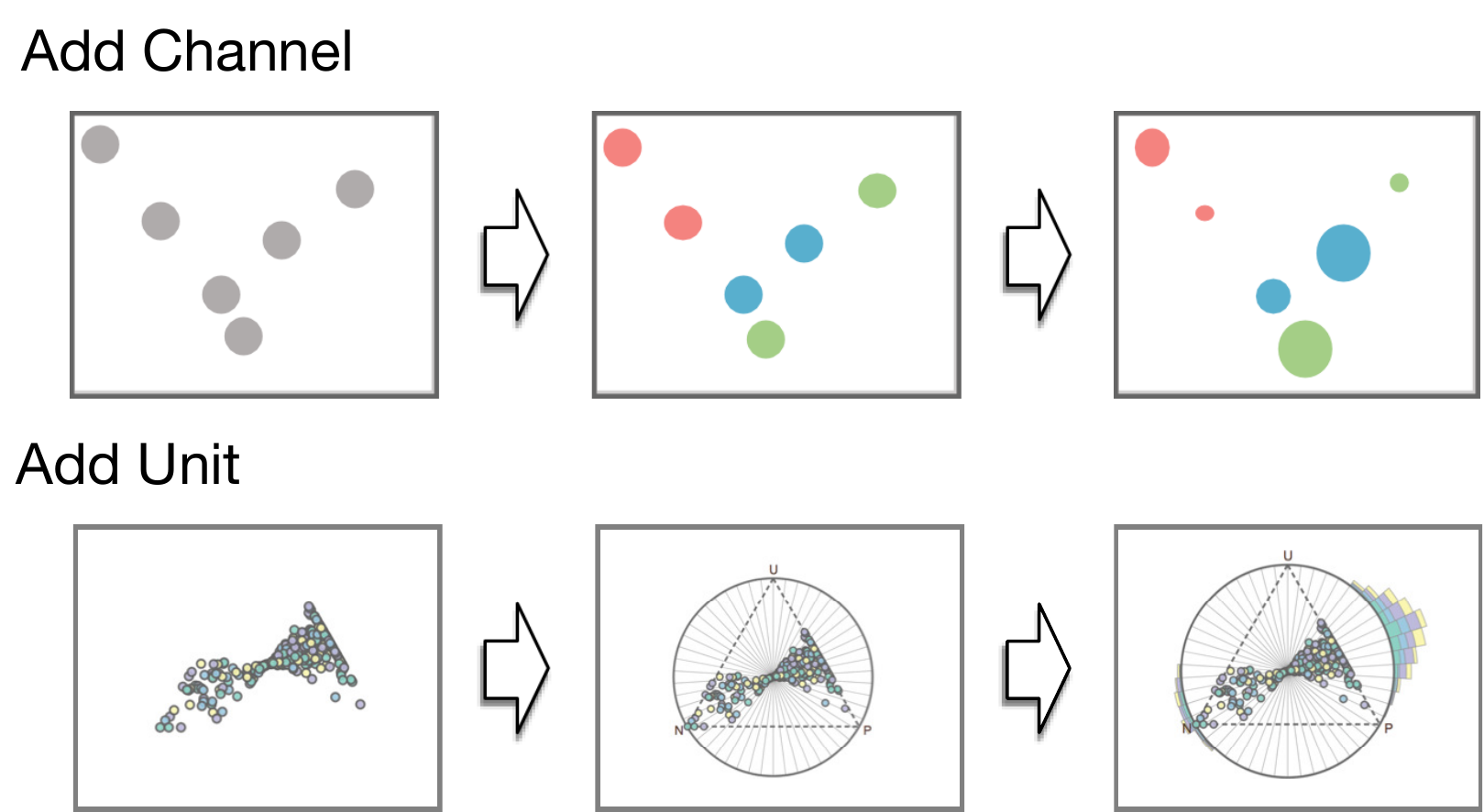}
\caption{Demonstration of the progressive introduction of visual channels (top) and visual units (bottom).
}
\label{fig:eg_template}
\end{figure}

\subsection{Phase2: Authoring}

In the Authoring Phase, teachers create an introduction slideshow by manipulating the visual units extracted in the Input Analysis Phase. 
In this section, we demonstrate the workflow of the Authoring Phase, 
which comprises two steps, namely, organizing units and introducing units (Fig.~\ref{fig:overview}). 


\subsubsection{Organizing Units}
The teachers define the relationships among the visual units,
and then organize a narrative sequence for introducing these units on the basis of the defined relationships.
In the generated slideshow tutorial, the visual units are added to the scene one by one following this sequence, 
preventing the addition of too much information at one step (R4). 

This step is conducted in \textbf{Units Panel}, which consists of two views, a node-link view and a sequence view (Fig.~\ref{fig:GOT}(c) and (d)).
In the node-link view, teachers define relationships between visual units, 
which can be independent (linked with a double arrow line), dependent (linked with an arrow line) or undefined (no link) (Section~\ref{subsubsec:rel_units}).
In the sequence view, Narvis suggests one narrative sequence 
on the basis of these relationships using a topological sort algorithm (R2).
Teachers are allowed to adjust this sequence as long as the adjustment doesn't conflict with the previously defined relationships.
This sequence can be inserted in the generated slideshow to inform the students of the overall structure (R6). 

\subsubsection{Introducing Units}

After determining the narrative sequence, teachers craft slides for the introduction of each unit.

Narvis allows teachers to add annotations and animated transitions for crafting a narrative introduction.
Annotation is a common and important technique used in narrative data visualization~\cite{mckenna2017visual, riche2018data}.
The effect of animated transition on improving perception and facilitating learning processes has been discussed by previous research~\cite{heer_animated_2007, ruchikachorn_learning_2015, Dessart_showing_2011}.

Two panels, \textit{i.e.,} \textbf{Channels Panel} and \textbf{Editing Panel}, are involved in this step.

\textbf{Channels Panel} (Fig.~\ref{fig:interface}(c)) displays possible channels for a unit, once it is selected in the Units Panel (Fig.~\ref{fig:interface}(b)).

These possible channels are detected by the analysis of the attributes of visual primitives in a visual unit.
For example, if visual primitives in a visual unit have different colors, color will be detected as a possible visual channel that needs to be explained.

We explicitly enumerate the possible visual channels, instead of requiring the teachers manually add them one by one. 
We believe that this mechanism can reduce, if not eliminate, the teachers' overlooking of some important visual encodings (R5).
By default, these channels are arranged in the decreasing order of visual salience.
Teachers can freely change the channels' order (\textit{e.g.}, explain simple encodings first), add and delete visual channels in the Channel Panel.
To explain a certain channel, teachers are required to add animated transitions and annotations.
Added animated transitions and annotations appear as tabs in the corresponding dotted box of the corresponding channel (Fig.~\ref{fig:interface}(c)).

\textbf{Editing Panel} displays an annotation or an animated transition (Fig.~\ref{fig:interface}(d)), once it is selected in the \textbf{Channels Panel}.
In \textbf{Editing Panel}, the teachers preview these annotations and animated transitions,
and then perform modifications such as resizing and moving symbol annotations, 
revising text annotations, 
and restricting an animated transition to specific primitives.

For an efficient authoring process, Narvis provides templates for adding annotations and animated transitions.
We propose these templates based on previous study~\cite{mckenna2017visual, riche2018data} and iterate over the design of these templates based on close discussions with three teachers mentioned in Section~\ref{sec:design_requirement} (T1, T3, T4).
Currently, Narvis supports seven types of animated transitions: fade-in, fade-out, growing, changing size, adding color, morphing, and highlighting, 
five types of symbol-based annotations: color legends, circles, arrow lines, double arrow lines, freeform lines,
and several text-based annotations for the explanation of different channels.
Fig.~\ref{fig:eg_template} demonstrates the progressive introduction of visual channels and visual units using the animated transitions provided by Narvis.
Teachers can also insert multiple/single choice questions in their slideshow tutorials by using the question annotation template.
These questions can remind the students of the previously mentioned information (R6).
Meanwhile, Narvis collects the student's answers to these questions, thereby helping the teachers evaluate their slideshows (R3).






\begin{figure}[tb]
\centering 
\includegraphics[width=\columnwidth]{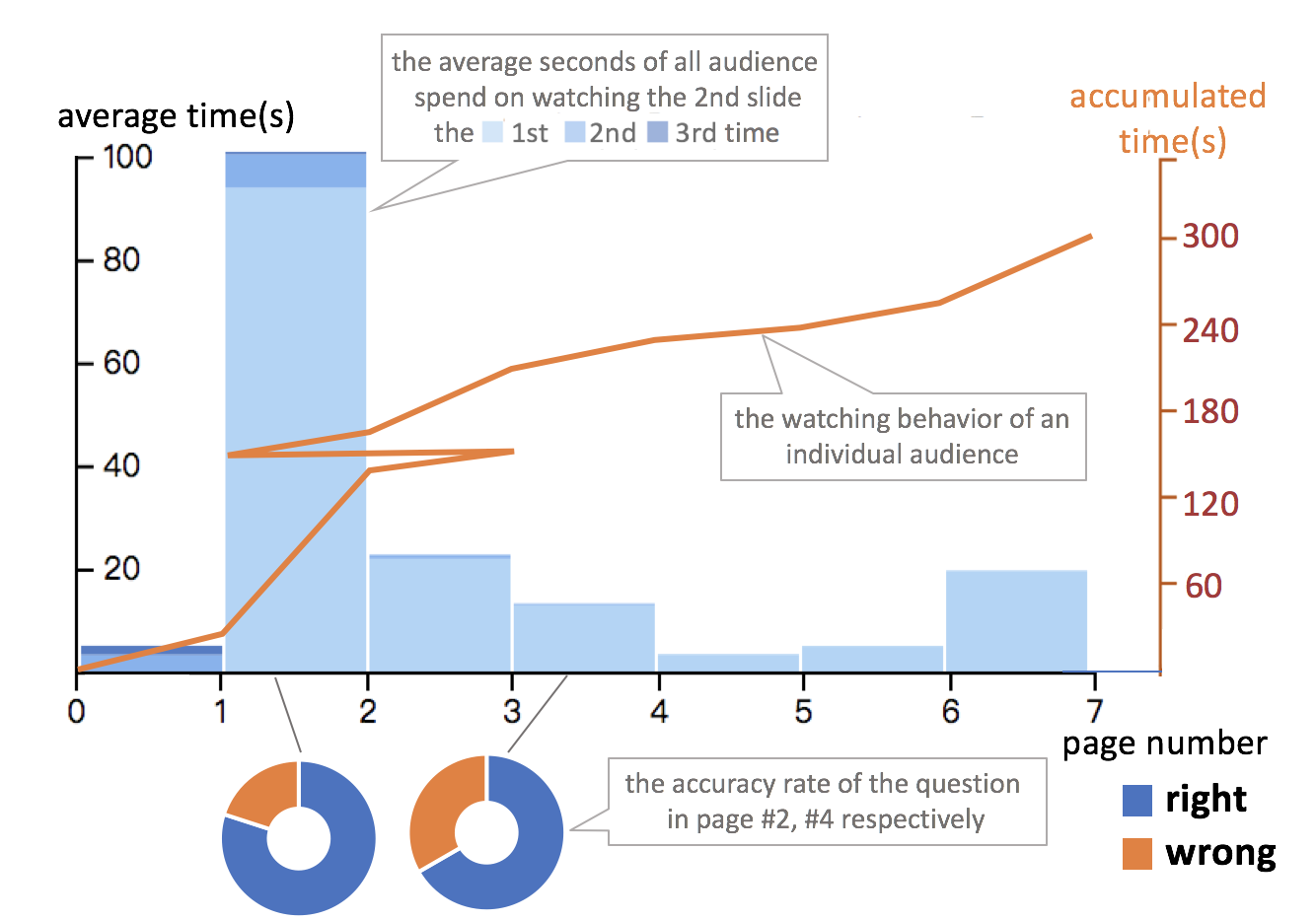}
\caption{Narvis visualizes the feedback data. 
The stacked bar chart describes the average watching time for different slides.
The line chart indicates the watching behavior of an individual student.
The donut chart represents the accuracy rate of student's answers if any question is inserted in the slideshow by the editor. }
\label{fig:feedback}
\end{figure}

\subsection{Phase3: Viewing}
The produced slideshow can be either saved locally as an HTML file, or saved in Narvis and watched online. 
For online viewing, Narvis collects the click activities of students 
(\textit{e.g.,} when they click a button to start a new slide or revert to a previous slide), 
their comments for this tutorial, 
and their answers to inserted questions, if any. 

The feedback data is visualized by Narvis in the form of a stacked bar chart, a line chart and donut charts (Fig.~\ref{fig:feedback}). 
The \textit{x}-axis represents the page number of slides in the bar chart and the line chart. 
In the line chart, the \textit{y}-axis represents the accumulated watching time of a particular student. 
By contrast, the \textit{y}-axis in the bar chart denotes the average watching time overall students for different slides. 
Each bar is split into colored bar segments. 
The bottom bar segment illustrates the time spent watching the slide the first time. 
If the student reverts and watches the slide for a second time, then a bar segment with a dark color is placed above the previous one, and so on. 
The rate of accuracy to each question is visualized as one donut chart linked with the slide containing the question.

The feedback data enables teachers to observe the students' behavior in watching the tutorial, 
check the students' understanding of the visualization design, thus generating ideas for later revisions (R3).

\begin{figure*}
\centering 
\includegraphics[width=\linewidth]{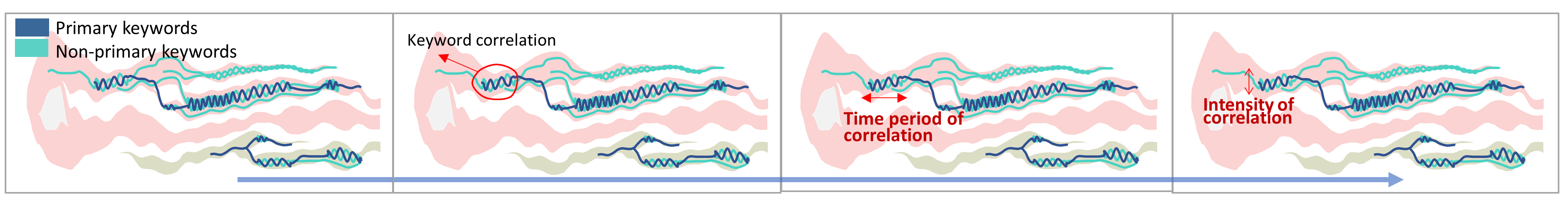}
\caption{Slides extracted from the slideshow created by PT3 with Narvis. 
These slides aim to explain two channels (\textit{i.e.,} color and size) of the ``lines'' in \textit{TextFlow} ~\cite{cui_textflow:_2011}}
\label{fig:user_study}
\end{figure*}  

\section{Evaluation}
In this section, we demonstrate the utility of Narvis through evaluating its authoring experience and the quality of the generated slideshows.

Narvis is tailored for creating introduction slideshows for visualization designs.
To the best of our knowledge, there is no other software that is specified for the same purpose.
A formal comparative study of Narvis with general presentation tools, such as Prezi and PowerPoint, would be biased,
since these tools lack the graphic editing capabilities that Narvis has.
Comparing Narvis with professional graphic editing tools, such as Adobe Illustrator and Adobe Photoshop, would also be unfair, 
since Narvis is tailored for introducing visual designs and provides more efficient operations.
Thus, we believe that it is more meaningful to show the utility of Narvis in qualitative studies.

\subsection{Study Design}
We conducted two sessions, \textit{i.e.}, Authoring Session and Viewing Session, 
to evaluate the authoring experience of Narvis and quality of the generated slideshows, respectively.

We recruited five participants as teachers (two females and three males) between 25 and 40 years old, denoted as PT1 to PT5, to produce introduction slideshows with Narvis.
The five teachers were all researchers or postgraduate students with considerable experience in data visualization.
We also sent emails to students in our university and recruited 20 volunteers (seven females and thirteen males) as students, 
denoted as PS1 to PS20, to evaluate the quality of the generated slideshow.
The twenty students were undergraduate or postgraduate students between 18 and 30 years old, with no prior background in data visualization.

\subsubsection{Material Preparation}

The authoring experience of Narvis is examined by asking five PTs to produce tutorial slideshows using Narvis.
We prepared a SVG file and a textual introduction in advance
so that teachers can better focus on producing tutorial slideshows. 
We chose \textit{TextFlow}\cite{cui_textflow:_2011} as the example for teachers to explain.
Two types of materials were prepared:
1) a textual description that explains the visual encodings of \textit{TextFlow}. 
This description is directly extracted from the original paper without any modification;
2) an SVG file generated by ourselves using D3.js.
We used the same data as used in \textit{TextFlow}, namely, 2980 articles related to ``Obama'' from January 25 to February 10 in Bing News\footnote{https://www.bing.com/news}.
The visualization is rendered as a web page.
We zoomed into topic flows with split/merge patterns and 
got the visualization as shown in Fig.~\ref{fig:user_study}.
We used the ``inspect elements'' function in Chrome DevTools\footnote{https://developers.google.com/web/tools/chrome-devtools/} to locate the SVG elements and the ``copy elements'' function to save them in a SVG file.

\subsubsection{Authoring Session}
The Authoring Session consisted of three phases: 
(1) \textbf{Sketching Phase}, 
(2) \textbf{Authoring Phase},
(3) \textbf{Feedback Phase}.
PTs were asked to think aloud during the whole session.
We were present in the room for the whole session to observe and take notes. 

In the \textbf{Sketching Phase}, 
PTs first learned a new visual design, \textit{TextFlow} proposed in ~\cite{cui_textflow:_2011},
through reading the literature description extracted from the paper. 
This learning phase cost about 20 minutes for each PT.
Then, they were asked to sketch ideas for introducing \textit{TextFlow}. 
They were required to enumerate all visual encodings and encouraged to consider 
(i) conveying insights to people with less experience in data visualization,
(ii) organizing a clear narrative structure,
and (iii) providing additional annotations and animated transitions. 
This phase took about 15 minutes for each PT.

In the \textbf{Authoring Phase}, 
PTs implemented the ideas in their sketches with Narvis.
We first demonstrated the workflow of Narvis and allowed PTs to familiarize
themselves with the Narvis using three basic visual designs, a chord diagram, a node-link diagram, and a parallel set.
During Authoring, PTs were asked to speak out 
i) ideas in their sketches that they failed or found inconvenient to implement with Narvis; 
ii) ideas that occurred to them while authoring with Narvis.
Examples of generated slideshows are demonstrated in Fig.~\ref{fig:teaser} and Fig.~\ref{fig:user_study}.

In the \textbf{Feedback Phase}, 
PTs first filled out a questionnaire, which evaluated Narvis on a five-point Likert scale.
Then, we asked these PTs open-ended questions to collect detailed comments and suggestions for Narvis.

\subsubsection{Viewing Session}
We ran a between-subject study in the Viewing Session.
Each PS watched one slideshow, which was randomly picked from the slideshows produced in Authoring Session. 
After watching, PSs were asked to finish a quiz, with a full mark of five, to check their understanding of the design of \textit{TextFlow}.
Then, they accomplished a questionnaire to rate the quality of the slideshow from one (very poor) to five (excellent) 
regarding readability (\textit{e.g.,} is it easy to read and follow the logic?), 
utility (\textit{e.g.,} does it help you understand this visual design?), 
aesthetics (\textit{e.g.,} does it look pretty and pleasant?)
and attractiveness (\textit{e.g.,} does it attract your interest?).
We asked them open-ended questions to collect detailed feedback and explanations for their rating.

\begin{table}[tb]
  \caption{Observation of five slideshows}
  \label{tab:slides}
  \small
  \centering
  \begin{tabu}{p{2cm}|p{0.9cm}|p{0.9cm}|p{0.9cm}|p{0.9cm}|p{0.9cm}}
  \toprule
 \textbf{} &\textbf{Slide 1} & \textbf{Slide 2} & \textbf{Slide 3}& \textbf{Slide 4} & \textbf{Slide 5} \\ 
   \midrule
  \textbf{Types of Animated Transitions} & Fade-in, Highlight & N/A & Fade-in, Highlight, Add-color & Fade-in, Highlight, Morphing & Fade-in, Add-color  \\ 
 \midrule
 \textbf{Number of Animated Transitions} & 2 & 0 & 4 & 4 & 3 \\ 
 \midrule
 \textbf{Number of Symbol-based Annotation} & 6 & 7 & 8 & 6 & 9 \\ 
 \midrule
 \textbf{Number of Text-based Annotation} & 6 & 2 & 8 & 6 & 3 \\ 
 \midrule
  \textbf{Reading Time (S) (mean, standard deviation)} & 127.05, 22.03 & 156.78, 14.08 & 169.33, 21.49 & 128.84, 19.43 & 143.79, 17.21\\ 
 \midrule
  \textbf{Information Missing   (in Slideshow /in Sketch) }& 1/2 & 0/1 & 0/2 & 0/0 & 0/0 \\ 
  \bottomrule
  \end{tabu}
  \vspace{1mm}
\end{table}

\begin{table}[tb]
  \caption{Overall evaluation of slideshows}
  \label{tab:rating}
  \small
  \centering
  \begin{tabu}{p{1.2cm}|p{1.2cm}|p{0.8cm}|p{0.9cm}|p{1.3cm}|p{0.9cm}}
  \toprule
 \textbf{} &\textbf{Readability} & \textbf{Utility} & \textbf{Aesthetics}& \textbf{Attractiveness} & \textbf{Quiz} \\ 
   \midrule
  \textbf{Mean} & 4.00 &  4.15 & 3.95 & 4.05 &4.475 \\ 
  \midrule
  \textbf{Standard Deviation} & 0.56 & 0.37 & 0.60 & 0.68  & 0.47\\ 
  \bottomrule
  \end{tabu}
  \vspace{1mm}
\end{table}

\subsection{Results}

Overall, participants are satisfied with the design of Narvis, regarding to the authoring experience of Narvis and the quality of the generated slideshows.
Participants in the teachers group commented that Narvis \textit{``is well-designed''}, \textit{``has clear UI''}, and \textit{``is easy to operate''}.
In the questionnaire, 
which was based on a 5-point Likert scale ranging
from strongly agree (5) to strongly disagree (1), they confirmed that the interaction is easy in general (mean 4, standard deviation 0) and they were able to craft a slideshow with Narvis after a short training period (4.8, 0.45).
They commented that the produced slideshows were visually pleasing (5, 0) and they will use Narvis in the future (4.4, 0.55).
We also evaluated the five slideshows generated in Authoring Session from the independent opinions of students.
Overall, 20 students were satisfied with the quality of the produced slideshows, as shown in Table ~\ref{tab:rating}.
The high scores they obtained in the subsequent quiz demonstrated that they were able to read the visualization design and obtain insights after watching the tutorials.

In the study, we have observed that the proposed design requirements (Section ~\ref{sec:design_requirement}) are fulfilled to some extent. For example,
\textbf{R1} and \textbf{R3} are met based on the results of the questionnaire and the quiz.
For \textbf{R1},
PTs agreed Narvis can implement the ideas they have sketched (mean 4.8, standard deviation 0.45) (\textit{i.e.,} expressive editing), reduce time and workload (4.8, 0.45) (\textit{i.e.,} efficient editing).
For \textbf{R3}, 
PTs appreciated the function of collecting feedback from the students,
agreeing that it would help them revise and improve the introduction slideshow (4, 0.71).

All PTs (strongly) agreed that Narvis helped them clarify their logic (4.8, 0.45) (\textbf{R5}), and offered a clear overview of the design (4.5, 0.5) (\textbf{R6}).
PSs also confirmed that the produced slides were easy to read (4, 0.56).
Four PTs (PT1, PT2, PT3, PT5) clearly expressed their appreciation for these designs in their detailed comments.
PT4 commented, \textit{``The Unit Panel helped me decide what should be introduced first and what should be introduced later.''}
PT5 commented, \textit{``This whole sequence organizer function is a refreshing idea. It alone can attract me to Narvis.''}

The fulfillment of \textbf{R4} is implicit, which is indicated by whether the produced slides are comprehensible.
PSs commented that the produced slides were easy to read (4, 0.56) and helped understand a visual design (4.15, 0.37).
The scores (4.475, 0.47) PSs got in the quiz also proved that they understood the introduced visualization design. 
We plan to exploit explicit measurement to evaluate the fulfillment of \textbf{R4}.

We examine the fulfillment of \textbf{R2}, avoid unconscious ignorance, based on our observation during Authoring Session.
For example, PT2 overlooked the encoding of glyph size in his sketch. 
He stopped with obvious hesitation when editing in the Channels Panel.
He then went back to check the provided textual description and added this missing encoding.
According to Table ~\ref{tab:slides},
failing to mention the encoding of certain channels when sketching occurred to three out of five teachers.
All the three teachers added several, if not all, missing channels during authoring with Narvis, which indicates that Narvis is able to remind teachers of the visual encoding they ignored.
Meanwhile, we also got negative feedback from two PTs about the restriction used in \textbf{R2}, indicating the need for further improvement about the designs related to \textbf{R2}.

\section{Discussion}
\subsection{Reflection on Evaluation Feedback}
We reflect on users' feedback on using Narvis and derive several design lessons, which can guide the further improvement of Narvis and inform other designs of 
visualization introduction.

First, the same type of users may have different, even opposite, preferences when editing information at different levels of details. 
To avoid information overload, Narvis allows only one visual channel and one visual unit to be explained at a time. 
While PTs highly appreciate the restricted explanation order of \emph{visual units}, they showed less interest in the restricted order of \emph{visual channels} and even complained that this restriction limited their editing.
A possible reason is that people prefer well-defined restrictions on high-level information (\textit{i.e., } visual units) and enjoy flexibility in editing detailed information (\textit{i.e.,} visual channels).

Second, different types of users may have different opinions for the same setting.
For example, participants in the students group were satisfied with the fact that only one visual channel was explained at one time.
However, some participants (PT2, PT4) in the teacher group thought such a restriction was too limited, commenting that they \textit{``might want to explain two channels at the same time''}.
This phenomenon indicates the importance of identifying different types of end users and understanding their different preferences.

\subsection{Encoding vs. Insight.}
When introducing a visualization design, 
the explanation of visual encodings and the introduciton of insights are sometimes mixed.
The mixed introduction of insights and encodings was also observed at three PTs (PT1, PT3, PT4) when using Narvis.
For example, after the annotation \textit{``height stands for the intensity of correlation''}, which explained the visual encoding, PT3 added another annotation \textit{``the keyword `reagan' has a weak correlation with keyword `debt' and a strong correlation with keyword `leader'''} to introduced an insight.

Narvis is designed to help users organize an explanation of visual encodings, 
which is the foundation to understand a visualization design and to obtain insights from it.
But Narvis also supports the introduction of insights through functions such as annotating and highlighting.
Users can introduce the patterns displayed by highlighting relevant components and adding annotations at appropriate time points (\textit{e.g.,} after introducing corresponding visual encodings).
We will explore how to offer more guidance for insight discovery in future work.

\subsection{Limitations}
Narvis has several limitations.

First, the usage context of the current version of Narvis may be limited.
From the perspective of end users, we merely interviewed students and teachers groups to derive design requirements.
In this scenario, students have strong motivation to learn the visualization design and understand visual encodings.
However, 
visualizations targeted at different end users may have different requirements for explanation.
For example,
readers in data journalism may care more about the insights a visualization conveys, and pay less attention to remembering visual encodings. Therefore, a pattern-based, or insight-based, explanation may be preferred.
In future research, we plan to generalize the application of Narvis by studying how visualization is explained and interpreted in various scenarios.
From the perspective of form, we only discussed tutorials in the form of slideshows, which are preferred by our interviewees for the explanation of complicated visualization designs.
However, other forms of tutorials can also be efficient in certain working scenarios.
For example, students in the interviews commented that legends can be helpful and efficient when the visual encodings are relatively simple.
It is beyond the scope of this work to identify suitable working scenarios for different forms of tutorials. 
In this paper, we assume that the explained visualizations are complicated enough and can hardly be explained by, for example, a simple legend.

Second, the evaluation lacks external validity.
We evaluated authoring experience and quality of the tutorials mainly based on self-reported perceptions.
As with all self-reported data, results of this evaluation have the potential for recalling bias, under-reporting, or over-reporting.
Meanwhile, whether the produced tutorials improve understanding of visualization designs is evaluated based on PSs' scores in a quiz where a baseline comparison is missing.
While the quiz results showed that participants in the students group correctly understood the visualization design, it is not clear the extent to which Narvis enhances the understanding of visualization designs.
Thus, current results should be treated with caution and we plan to investigate whether our insights can apply more generally in future work. 
Nevertheless, we are encouraged by the fact that the current version of Narvis was appreciated by the initial users and got positive feedback.

Third, the input analysis method we implemented can be further improved.
Currently, Narvis processes the input visualization as an SVG file.
The components in the original visualization that are rendered as HTML elements (\textit{e.g.,} tooltip as a $<div>$) are lost during the input analysis. 
Meanwhile, the analysis results are influenced by the structure of the input SVG. 
More specifically, if the input SVG has nested groups, the tree list will be complex and require more efforts from the teachers to select visual units.
These problems can be alleviated by using more sophisticated input analysis methods.

\section{Conclusion and Future Work}
In this paper, we present Narvis, an authoring tool to generate slideshows for explaining visualization designs. 
The design and implementation of Narvis are guided by our understanding of two types of end users, namely, teachers and students.
Narvis provides a sequence organizer for clear narrative structures and
a series of templates for easy implementation of common operations. 
Moreover, Narvis offers mechanisms for avoiding information overload and unconscious information omission. 
Thus, using Narvis results in an efficient crafting process and a high-quality generated slideshow.
User studies have confirmed the utility and effectiveness of Narvis. 
To the best of our knowledge, this study is the first presentation tool tailored for the introduction of visual designs.


We envision improving Narvis in several directions.
First, 
we will improve the design of Channel Panel to allow users more freedom of editing and to provide editing guidance without undermining flexibility. 
Second, 
we plan to better support the introduction of insights in Narvis.
A possible solution is to automatically detect patterns (\textit{e.g., }outliers, clusters, extremes) and give hints to introduce insights after the explanation of corresponding visual channels.
Third, 
we are interested in analyzing the differences between data visualizations targeted at different end users (\textit{e.g.,} readers of data journalism, subject specialists) and their requirements of explaining.
By doing so, we aim to extend the usage context of Narvis.

\acknowledgments{
The authors would like to thank all the participants involved in the studies and the reviewers for their valuable feedback.
This work is partially supported by funding from the Theme-based Research Scheme of the Hong Kong Research Grants Council, project number T44-707/16-N.
}

\bibliographystyle{abbrv-doi}

\bibliography{template}
\end{document}